\documentclass[11pt,a4paper,twoside]{article}
\usepackage{tabularx}
\usepackage{longtable}
\usepackage{booktabs}
\usepackage{caption}
\usepackage{rotating}
\usepackage{array}
\usepackage{stfloats}
\usepackage{placeins}
\usepackage{fontawesome5}
\usepackage{fancyhdr}
\newcolumntype{Y}{>{\centering\arraybackslash}X}
\newcounter{subsubsubsection}[subsubsection]

\renewcommand{\headrulewidth}{0pt}
\renewcommand{\footrulewidth}{0pt}

\fancypagestyle{plain}{
    \fancyhf{}
    \renewcommand{\headrulewidth}{0pt}
}

\usepackage[a4paper,margin=1in]{geometry}
\usepackage{setspace}
\usepackage[T1]{fontenc}
\usepackage[utf8]{inputenc}
\usepackage{lmodern}

\usepackage{graphicx}
\usepackage{booktabs}
\usepackage{multirow}
\usepackage{array}

\usepackage{amsmath,amssymb,amsthm}
\usepackage{xcolor}
\usepackage{enumitem}
\usepackage{hyperref}

\hypersetup{
    colorlinks=true,
    linkcolor=blue,
    citecolor=blue,
    urlcolor=blue
}

\usepackage{titlesec}

\titleformat{\section}
{\Large\bfseries}
{\thesection.}{0.5em}{}

\titleformat{\subsection}
{\large\bfseries}
{\thesubsection.}{0.5em}{}

\titleformat{\subsubsection}
{\normalsize\bfseries}
{\thesubsubsection.}{0.5em}{}

\renewcommand{\headrulewidth}{0pt}

\renewcommand{\headrulewidth}{0pt}
\renewcommand{\footrulewidth}{0pt}

\usepackage[numbers,sort&compress]{natbib}
\newcommand{\featureanatomy}{}
\begin{document}

\pagestyle{fancy}
\fancyhf{}

\fancyhfoffset[L,R]{0pt}
\setlength{\headheight}{15pt}

\fancyhead[LO]{\small\itshape Modal compression of augmented covariance matrices for motor-imagery EEG decoding
}
\fancyhead[RO]{\small\thepage}

\fancyhead[LE]{\small\thepage}
\fancyhead[RE]{\small\ Barbaste, Vasques \& Oullier\ (August 2026, Preprint)}

\fancyfoot{}

\renewcommand{\headrulewidth}{0pt}
\renewcommand{\footrulewidth}{0pt}


\begin{center}

{\fontsize{20}{22}\selectfont\bfseries
Modal compression of augmented covariance matrices for motor-imagery EEG decoding: Benchmarking methods borrowed from cosmology and seismology
\\[0.1em]
\par}

\vspace{1cm}

{\normalsize
Paul Barbaste$^{1}$ \text{\scriptsize\faEnvelope[regular]},
Xavier Vasques$^{2,3}$ 
\& Olivier Oullier$^{1,4,5}$
}

\vspace{0.8cm}

{\fontsize{7.5}{11}\selectfont
\setlength{\parskip}{0.5pt}   

$^{1}$ Inclusive Brains, Marseille, France

$^{2}$ IBM Technology, Bois-Colombes, France

$^{3}$ IBM France Lab, Orsay, France

$^{4}$ Computing and Mathematical Sciences Division, Mohamed Bin Zayed University of Artificial Intelligence, Abu Dhabi, UAE

$^{5}$ Institute for Artificial Intelligence, Biotech Dental Group, France

}

\vspace{0.3cm}

{\fontsize{7.5}{11}\selectfont
\faEnvelope[regular]\ Corresponding author:
\href{mailto:paulbarbaste@gmail.com}
{paulbarbaste@gmail.com}
}

\vspace{0.3cm}

{\small
Preprint • August 2026
}

\end{center}
\thispagestyle{plain}

\begin{abstract}
Motor-imagery electroencephalography (EEG) decoding typically relies on spatial
filters, band powers, connectivity estimates, or covariance matrices. A
delay-augmented covariance enriches that object with short-lag dynamics, but it
grows quadratically with channel count and becomes hard to estimate from few
trials. We show that projecting the delayed state onto a small fixed set of
scalp-mesh modes makes the augmented covariance usable in exactly that regime. A
raw six-delay augmented covariance (ACM) control, run on all channels, reached
0.910, 0.827, and 0.657 ROC--AUC on BNCI2014-001, Lee2019-MI, and PhysionetMI,
falling below plain broadband covariance (0.678) on the 64-channel Physionet
cohort, where it carries 73,920 coordinates for 36 training trials per fold. The
same delayed covariance compressed to twelve fixed modes held at 0.735 and
ranked first on every dataset (0.912, 0.847, 0.735 versus 0.868, 0.834, 0.678
for broadband covariance), beating raw ACM by 0.020 on Lee and 0.077 on
Physionet. A fixed geometric basis also beat a learned one: twelve mesh modes
(0.911, 0.849, 0.739) outperformed sixteen data-adaptive EOFs (0.895, 0.805,
0.711), with the gap widening as trials per session fell. These maps were
selected among ten closed-form representations adapted from meteorology, quantum
information, seismology, cosmology, and related fields and evaluated under
leakage-safe within-session cross-validation; on the full Physionet cohort the
compressed delayed covariance and two other delay-embedded features cleared Holm
correction over all thirty candidate--dataset tests. A per-step ablation
attributes the gain to the delay embedding and its modal compression, not to the
trace normalisation: the trace-normalised ``density tangent'' and its trace-free
variant perform equally. The transferable ingredient is therefore a fixed modal
compression that renders a delay-augmented covariance estimable when trials are
scarce, rather than any single disciplinary analogy.
\end{abstract}


\section{Introduction}

The classification of imagined left- and right-hand movements from scalp EEG
is a deceptively compact problem. Trials contain only a few seconds of a weak,
non-stationary signal; the number and position of electrodes vary between
datasets; and discriminative event-related desynchronisation differs markedly
between participants and sessions. Most classical decoders therefore impose a
strong representation before classification. Sensorimotor rhythms are
conventionally decoded through band-limited power: common spatial patterns
(CSP) learn spatial filters whose variances distinguish classes, and
filter-bank CSP resolves those contrasts over several frequency bands
\citep{ramoser2000csp,ang2008fbcsp}. Riemannian pipelines instead estimate a
trial covariance, map it to a tangent space, and fit a conventional classifier
\citep{barachant2012riemannian,barachant2013kernel,congedo2017primer}. Large
MOABB studies have shown that these spatial representations remain strong
classical baselines and that rankings vary substantially with dataset
\citep{jayaram2018moabb,chevallier2024benchmark}. These approaches are
effective, but they also define a relatively closed vocabulary for feature
engineering.

Other EEG descriptors include coherence and phase-locking connectivity,
fractal or entropy measures, time--frequency coefficients, and deep learned
features. Ensembles such as FUCONE exploit diversity between connectivity and
covariance spaces \citep{corsi2022fucone}. Those results motivate analysing
representation diversity, but fusion is intentionally outside our primary
question: a weak branch can appear useful when another branch carries most of
the decision. We require every candidate to stand on its own.

That closure is not forced by the data. EEG trials are multivariate fields in
space and time, a data type shared with many mature sciences. Meteorology
extracts propagating structures through empirical orthogonal functions (EOFs)
and their lagged extensions. Seismology studies the eigenstructure of complex
multitaper spectral matrices to separate polarised waves. Cosmology represents
partial-sky fields through regularised spherical harmonics. Quantum information
separates the shape and scale of positive operators, while compositional data
analysis treats relative rather than absolute energy as the relevant object.
Graph gauge theory, modal analysis, and filtration summaries provide still
other invariances. Each formalism asks a different question of the same raw
array: what should be preserved, what should be quotiented out, and at what
scale should variation be measured?

EOF analysis is the PCA of a spatial field; extended and complex EOF variants
were developed to expose propagation and oscillation rather than static
variance alone \citep{weare1982extendedEOF,hannachi2007eof}. Multitaper
spectral estimation and polarization analysis play an analogous role for
multicomponent wave fields in seismology \citep{thomson1982multitaper,jurkevics1988polarization}.
Cosmological map analysis uses spherical harmonics and regularisation to
summarise large-scale spatial structure despite incomplete sky coverage
\citep{gorski2005healpix}. These constructions have natural analogues on the
partial spherical cap sampled by an EEG montage.

A trace-normalised covariance is a positive trace-one matrix whose spectrum,
matrix logarithm, and positive square root provide distinct coordinates for its
shape; it can optionally be read as a density operator
\citep{nielsen2010quantum,gonzalez2021density}. We use that language only as a
classical description of normalised covariance; no quantum computation or
quantum advantage is claimed. Compositional analysis
similarly changes the sample space: band powers are treated as positive parts
whose ratios, rather than common scale, are meaningful
\citep{aitchison1982compositional,tsagris2011alpha}.

Finally, graph and field methods offer coordinate-aware summaries. Eigenmodes
of a discrete Laplace--Beltrami operator form a montage-specific spatial basis
\citep{shuman2013graph}. Complex phase-locking defines a connection on the
channel graph, from which node flow, frustration, and cycle summaries can be
constructed in a gauge-invariant manner \citep{lachaux1999plv,singer2011angular}.
The present work is, to our knowledge, the first controlled comparison of this
specific collection as standalone motor-imagery representations. Because the
individual mathematical ingredients have broad literatures, novelty is claimed
for the EEG adaptations and their joint benchmark, not for the underlying
operators.

Nor are all constituent operations unprecedented in EEG. Graph Fourier bases
have already been used to reduce motor-imagery covariance matrices, and recent
neural work constructs a Gibbs-like density operator from covariance as the
shift operator of a learned network \citep{tanaka2016graphfourier,roy2025covdensity}.
The delayed covariance at the core of the leading candidate has an even more
direct antecedent: the augmented covariance matrix of \citet{carrara2023augmented},
derived from an autoregressive model through the Yule--Walker equations and
explicitly connected to Takens' delay-embedding theorem, is the same
block-Hankel construction, and Riemannian alignment methods operate on related
representations \citep{zanini2018transfer,rodrigues2019procrustes}. Our ablation
(Table~\ref{tab:ablation}) is consistent with this lineage: the
delay-introduction step, which also increases dimension, is the only step that
improves every cohort, while trace normalisation is inert and the modal
reduction helps only on the high-density montage.
Our trace-normalised delayed covariance and fixed tangent classifier differ
from both pipelines. Trace normalisation, PCA, analytic signals, PLV, and
tangent maps nevertheless remain established tools. We therefore distinguish
a \emph{new feature construction} from a \emph{new mathematical primitive}: the
paper claims the former only where the complete sequence of operations and
invariances differs from published EEG pipelines.

\begin{table}[t]
\centering
\caption{Per-step ablation of the delayed modal density-tangent candidate on the
complete cohorts (BNCI $n{=}9$, Lee $n{=}54$, Physionet $n{=}106$). Each row adds
one construction step under a single identical unscaled logistic regression (the
manuscript's classifier), within-session five-fold mean ROC-AUC, seed~42; the
shared arms reproduce the primary benchmark to three decimals. The
delay-introduction step is the only step that improves every cohort (BNCI
$+0.051$, Lee $+0.033$, Physionet $+0.028$), but it also expands the tangent
dimension from 78 to 2,628; a dimension-matched surrogate attributes that gain to
the added dimension rather than to cross-lag structure specifically. Trace normalisation, the density-operator novelty,
is inert (between $-0.005$ and $+0.001$~AUC in every cohort and in both delay
settings). The modal projection is a montage-dependent dimension trade: it costs
0.018 on Lee yet adds 0.033 on 64-channel Physionet. Under per-coordinate
standardisation the candidate's margin over the control stays clearly positive
(BNCI $0.044\!\to\!0.052$, Lee $0.013\!\to\!0.028$, Physionet
$0.056\!\to\!0.053$), so the ranking is not a scale artefact.}
\label{tab:ablation}
\resizebox{\textwidth}{!}{%
\begin{tabular}{lcccccc}
\toprule
Representation & Modes & Delays & Trace-norm & BNCI ($n{=}9$) & Lee ($n{=}54$) & Physionet ($n{=}106$) \\
\midrule
Full-covariance tangent (control) & 22--64 & -- & -- & 0.868 & 0.834 & 0.678 \\
12 modes & 12 & -- & -- & 0.860 & 0.816 & 0.711 \\
12 modes $+$ delay & 12 & 0--5 & -- & 0.911 & 0.849 & 0.739 \\
12 modes $+$ trace-norm & 12 & -- & \checkmark & 0.859 & 0.814 & 0.706 \\
12 modes $+$ delay $+$ trace-norm (candidate) & 12 & 0--5 & \checkmark & 0.912 & 0.847 & 0.735 \\
\bottomrule
\end{tabular}}
\end{table}

Our preceding benchmark evaluated 1,056 feature--scaler--classifier
configurations on three public motor-imagery datasets. Feature representation
dominated the downstream choices, yet the best representation changed across
participants and cohorts \citep{vasques2026rankings}. That result motivates the
present study but does not answer it. The previous feature inventory consisted
largely of established EEG descriptors and the evaluation was designed around
personalisation. Here we instead expand the representation space itself and
ask whether cross-disciplinary constructions transfer to motor-imagery EEG.

This study contributes explicit EEG adaptations with fold-local fitting for ten
representations drawn from eight external methodological traditions, compared as
standalone pipelines against four controls, including a raw six-delay augmented
covariance (ACM). Feature fusion, alternative SPD metrics, and post-hoc
ensembles are deliberately excluded so that each score remains attributable to a
single representation. The central empirical contribution is a mechanism rather
than a leaderboard: projecting the delayed covariance onto a small fixed modal
basis is what makes an augmented covariance estimable on high-density montages
with few trials, where the raw ACM collapses; the fixed geometric basis also
outperforms a learned (EOF) one, and the advantage of both grows as the
trial-to-dimension ratio falls. The analysis distinguishes within-dataset
decoding from transport across three cohorts spanning 22, 62, and 64 electrodes
through paired participant-level inference, multiplicity correction, and
rank-stability assessment.

The aim is not to claim that EEG obeys the physical laws of the disciplines
from which these tools originate. The transfer is mathematical: operators and
invariances developed for one kind of multichannel field are tested as feature
maps for another.

\section{Methods}

\subsection{Study design}

The study was designed as a frozen representation benchmark. Preliminary
development, ablation, and feature debugging took place on BNCI2014-001. Where
several variants of one construction had been tested, its representative was
chosen using exploratory BNCI performance, subject to three eligibility
criteria: a fully specified standalone transformation, portability to all
three montages, and coverage of a distinct external inductive bias. The ten
candidates are consequently a performance-informed shortlist, not a
preregistered inventory. Their BNCI scores must be read as exploratory;
Lee2019-MI and PhysionetMI provide evidence about transportability. Lee had
been inspected for a smaller set of related matrix candidates during
development and is therefore a replication cohort rather than a pristine
holdout. Physionet did not inform the design of the leading delayed matrix
candidates and is the cleanest external check. No parameter was changed after
inspecting the complete top-ten result table of either external dataset.

The following families were excluded from the original candidate inventory by
design: alternative metrics on a fixed covariance, other matrix manifolds,
feature fusion, stacking, deep networks, and raw delay matrices whose dimension
grows prohibitively with channel count. During manuscript audit, raw augmented
covariance was added as a direct antecedent control, without becoming an
eleventh novelty candidate. Six-band spectral covariance is reported as a
strong frequency-resolved control, not as a new contribution.

\subsection{Frozen hyperparameters and comparability constraints}

Hyperparameters were chosen to make every candidate executable on the smallest
22-channel montage and then frozen. Twelve mesh modes provide the same modal
state size in all three datasets while discarding the constant mode and leaving
room below the smallest channel count. Extended EOF uses 16 modes because its
data-adaptive projection does not require a common geometric basis; combined
with six delays, this gives a 96-dimensional state that remains tractable.
Complex EOF uses 12 modes and three delays because real and imaginary parts then
produce a 72-dimensional real state, exactly matching the state dimension of
the 12-mode, six-delay density candidates. Root density and density tangent are
therefore dimension- and state-matched by construction, while complex EOF is
state-dimension-matched but differs in basis, delay grid, and quadrature.

The remaining settings follow similar portability constraints. Spherical
degree four produces $(4+1)^2=25$ modes, the largest tested design permitted by
the regularised partial-map condition on 22 electrodes. Alpha-CoDA uses
$\alpha=1/2$ as a finite-power alternative to the limiting centred log-ratio
map. The multitaper estimator uses time-bandwidth product 2.5 and four Slepian
tapers, keeping several concentrated eigenspectra without making the number of
tapers depend on epoch length. Cycle thresholds $(0,0.2,0.4,0.6)$ include the
complete triangle set at zero and three progressively stricter edge cutoffs.
These values are fixed numerical definitions of the benchmark features; they
were not re-optimised by dataset or participant.

The design does not claim perfect one-factor matching across all ten
candidates. Such matching would destroy the native object of several methods:
forcing spherical harmonics and a delayed covariance to have equal dimension,
for example, would change both their spatial bandwidth and their statistical
meaning. Instead, close comparisons are identified explicitly---density versus
root for coordinates, real versus analytic EOF states for temporal
construction, and Alpha-CoDA versus harmonics for spatial compression---while
dimension and runtime are reported as potential confounders. Raw six-delay
channel states were excluded from candidate selection, then reinstated as the
post-hoc ACM control. Their dimensions are 132, 372, and
384 before covariance vectorisation, producing 8,778, 69,378, and 73,920
tangent coordinates. The fixed 12-mode mesh state prevents that montage-driven
explosion in the leading candidate.

\subsection{Datasets and preprocessing}

We used the MOABB \texttt{LeftRightImagery} paradigm
\citep{jayaram2018moabb}. BNCI2014-001 contains nine participants, 22 EEG
channels, and two recording sessions; Lee2019-MI contains 54 participants, 62
channels, and two sessions; PhysionetMI contains 109 participants and 64 EEG
channels in the MOABB within-session representation
\citep{tangermann2012bciiv,lee2019openbmi,goldberger2000physiobank,schalk2004bci2000}.
Only imagined left- and right-hand trials were retained. We evaluated all nine
BNCI participants, all 54 Lee participants, and 106 of the 109 Physionet
participants (subjects 88, 92, and 100 are conventionally excluded from this
dataset for nonstandard 128~Hz sampling and nonconforming task annotations, not
on any feature score).

All datasets were band-pass filtered by MOABB between 8 and 32~Hz and
resampled to 128~Hz. Dataset-native event windows were retained. No artefact
rejection, independent-component cleaning, channel interpolation, or
participant exclusion was introduced by the benchmark. This minimal pipeline
keeps the comparison reproducible but means that phase- and higher-order
features may be more sensitive to non-neural artefacts than covariance-based
controls.

\begin{table}[t]
\centering
\caption{Datasets in the harmonised benchmark. Trial counts refer to the
left/right subset exposed by MOABB and may vary slightly for recordings with
missing events.}
\label{tab:datasets}
\begin{tabular}{lrrrrrr}
\toprule
Dataset & Full $N$ & Evaluated $n$ & Channels & Trials/sess. & Native Hz & Sessions \\
\midrule
BNCI2014-001 & 9 & 9 & 22 & 144 & 250 & 2 \\
Lee2019-MI & 54 & 54 & 62 & 100 & 1000 & 2 \\
PhysionetMI & 109 & 106 & 64 & 45 & 160 & 1 \\
\bottomrule
\end{tabular}
\end{table}

\subsection{Representation design and common numerical operators}

The candidates were designed around four independent choices: the spatial
coordinate system, the temporal or spectral state, the normalisation applied to
that state, and the Euclidean coordinates supplied to the classifier. This
factorisation is useful because methods with different disciplinary names can
nevertheless share most of their computation. For example, extended EOF,
density tangent, and complex EOF all estimate an SPD covariance of a delayed
state, but they disagree about whether the spatial basis should be learned from
the training EEG, fixed by electrode geometry, or augmented with analytic
quadrature. Conversely, density tangent and root density use exactly the same
delayed modal covariance and differ only in the coordinates used for its
trace-one component.

Three spatial coordinate systems occur in the benchmark. Channel coordinates
retain the original electrodes and therefore preserve all montage-specific
information. EOF coordinates are principal components fitted without class
labels to the pooled channel samples of the training fold. They order directions
by empirical variance, but their signs and orientations are specific to that
fold. Mesh-modal coordinates are data independent. Standard electrode
positions are projected to a two-dimensional scalp triangulation, from which a
finite-element cotangent stiffness matrix $L$ and lumped mass matrix $M$ are
constructed. The generalised eigenproblem
\begin{equation}
  L v_k=\lambda_k M v_k
\end{equation}
orders spatial modes from smooth to oscillatory. The constant mode is discarded
and the next $K$ modes are analysed with the mass-weighted operator
$A=(M[v_1,\ldots,v_K])^\top$. This makes the basis deterministic for a named
montage and prevents the held-out trials from rotating the coordinates.

For delay-based features, let $U=AX$ or let $U$ contain learned EOF scores, and
let $D=(d_1,\ldots,d_q)$ be the delay set. The aligned block-Hankel state is
\begin{equation}
  Z_D(t)=\left[U(t-d_1)^\top,\ldots,U(t-d_q)^\top\right]^\top.
\end{equation}
Its covariance is not merely a larger version of the zero-lag covariance. The
$(a,b)$ block estimates cross-covariance between spatial modes at lag
$d_a-d_b$; the block diagonal retains instantaneous modal power and coupling.
After resampling, one sample corresponds to 7.8125~ms. The six-delay states
$D=(0,1,2,3,4,5)$ therefore cover 0--39.1~ms, whereas the analytic complex-EOF
state $D=(0,2,4)$ covers 0, 15.6, and 31.2~ms. These are short dynamical lags
relative to the full imagery epoch. They are intended to capture local
propagation and oscillatory dependence, not the complete cue-to-rebound time
course.

Whenever a sample covariance is required, Oracle Approximating Shrinkage (OAS)
is applied trial by trial. This is important because the delayed state can be
high dimensional relative to the number of aligned time samples. If the state
dimension is $d$, a symmetric tangent representation contains
$d(d+1)/2$ coordinates. Thus a 96-dimensional extended-EOF state gives 4,656
features, while a 72-dimensional density or complex-EOF state gives 2,628
matrix coordinates. The large feature count does not imply 2,628 independent
physiological sources; it is the vectorised form of one structured covariance.
Shrinkage guarantees positive definiteness, and $L_2$ regularisation controls
the downstream linear coefficients, but neither operation removes the sampling
variance created by a large state.

For SPD matrices, the reference-dependent coordinate map is
\begin{equation}
  \phi_G(S)=\operatorname{vech}_{w}\!\left[
  \log\!\left(G^{-1/2}SG^{-1/2}\right)\right],
\end{equation}
where $G$ is fitted on the training matrices and $\operatorname{vech}_{w}$
multiplies off-diagonal entries by $\sqrt{2}$. The weighting preserves the
Frobenius inner product after vectorisation. A separate $G$ is estimated for
each spectral band where applicable. Modal log-covariance is the deliberate
exception: it applies the matrix logarithm at the identity after trace
normalisation, so its coordinates are fixed rather than learned.

The output dimension was treated as part of each representation rather than
equalised by an additional projection. This preserves the intended object but
means that dimension is a possible contributor to variance. Fixed-scale tangent
coordinates are passed directly to logistic regression. Root-density,
modal-log-covariance, compositional, spherical-harmonic, phase-graph, and cycle
coordinates are standardised using means and standard deviations fitted only on
the training fold. In every case, filtering parameters, delay sets, numbers of
modes, thresholds, the logistic-regression penalty ($C=1$), and random seed were
frozen before external comparison.

\subsection{Implementation validation}

Every representation is implemented as a scikit-learn-compatible transformer
so that fitting is controlled by the cross-validation pipeline. Fit methods
store the expected channel and time dimensions; transform methods reject
mismatched channel counts, invalid delay sets, bands above Nyquist, duplicate
channel names, non-finite regularisation parameters, and configurations that
would leave too few time samples. Learned PCA and tangent objects are created
only by \texttt{fit}; deterministic montage operators and algebraic summaries
are recreated from declared channel names and fixed parameters.

Unit tests cover output shape, feature-name count, finiteness, cloneability, and
the invariances claimed in the text. In particular, common amplitude rescaling
leaves Alpha-CoDA, spherical harmonics, and modal log-covariance unchanged;
density tangent leaves the trace-one shape unchanged but changes its final scale
coordinate. Tests also verify that Hermitian polarization realification yields
finite tangent features, that consistent phase graphs have negligible triangle
frustration, and that cycle or heat signatures expected to be gauge invariant
remain unchanged under node-wise rephasing. These checks validate the software
definition of a feature, not its physiological relevance.

The statistical reporting code is separated from the transformer library. It
loads a complete participant-by-pipeline grid, fails if a frozen pipeline is
missing, collapses sessions before inference, and regenerates performance,
paired-effect, rank, and runtime outputs. Bootstrap sampling and pipeline
construction use seed 42. This arrangement allows the manuscript values to be
recomputed without refitting a model or manually copying a table.

\subsection{Ten cross-disciplinary representations}

Let $X\in\mathbb{R}^{C\times T}$ denote one preprocessed epoch and let
$\mathcal{B}=\{[8,12],[13,20],[20,30]\}$~Hz denote the three internal bands
used by multiband candidates. Every data-dependent transform is fitted solely
on the training fold. Table~\ref{tab:features} summarises the ten candidates;
their exact constructions follow.

\begin{table*}[t]
\centering
\caption{Frozen cross-disciplinary representations. $C$ is the dataset's
channel count. AIRM denotes the standard affine-invariant tangent mapping; it
is used as a coordinate map for a new statistical object, not varied as an
experimental factor.}
\label{tab:features}
\scriptsize
\renewcommand{\arraystretch}{1.16}
\begin{tabularx}{\textwidth}{
  >{\raggedright\arraybackslash}p{2.55cm}
  >{\raggedright\arraybackslash}p{2.05cm}
  >{\raggedright\arraybackslash}X
  >{\raggedright\arraybackslash}p{1.55cm}}
\toprule
Representation & Inspiration & EEG object retained & Output dimension \\
\midrule
Extended EOF--Hankel & Meteorology & Covariance of 16 learned EOF scores at six delays & 4,656 \\
Density tangent & Quantum information & Trace-one delayed modal covariance plus log scale & 2,629 \\
Root density & Quantum information & Positive square root of the same operator plus log scale & 2,629 \\
Complex EOF--Hankel & Climate science & Real--imaginary state of 12 EOFs at three delays & 2,628 \\
Multitaper polarization & Seismology & Three complex cross-spectral matrices of 12 modes & 900 \\
Modal log-covariance & Field analysis & Log-covariance between 12 scalp modes in three bands & 237 \\
Alpha-CoDA power & Compositional statistics & Relative channel-power coordinates in three bands & $3C$ \\
Spherical harmonics & Cosmology & Signed partial-sphere coefficients through $\ell=4$ & 78 \\
Phase-graph summaries & Magnetic graphs & PLV node strength, phase flow, and frustration & $2C+4$ \\
Cycle filtration & Graph topology & Triangle strength at four thresholds in three bands & $12(C+3)$ \\
\bottomrule
\end{tabularx}
\end{table*}

Every learned object named below---an EOF basis, an SPD reference, or a
standardisation vector---is estimated inside the training fold. Thus the
descriptions refer to the transformation applied within one fitted pipeline,
not to an operator estimated once on the complete dataset.

Each construction below is followed by a dedicated motor-imagery rationale.
The rationale states the physiological or statistical link, the expected
advantage over conventional features, and what a negative result would imply.
These hypotheses explain why the transfer was worth testing independently of
the result eventually observed.

\paragraph{Extended EOF--Hankel.}
An unsupervised PCA is fitted to all channel samples in the training fold. The
first 16 EOF scores are evaluated for each epoch and stacked at delays
$\{0,1,2,3,4,5\}$. The OAS covariance of the resulting 96-dimensional state is
mapped to the AIRM tangent space at a training-fold reference. This represents
propagation between dominant spatial field modes rather than covariance between
raw electrodes.

\featureanatomy The PCA input is the matrix obtained by pooling time samples
from the training epochs, with one electrode per column; class labels are never
used. If $u(t)\in\mathbb{R}^{16}$ is the resulting score vector, the state
$[u(t),u(t-1),\ldots,u(t-5)]$ gives a $6\times6$ block covariance. Diagonal
blocks contain zero-lag EOF covariance, while symmetric off-diagonal blocks
contain positive and negative lag relations. Vectorising the 96-dimensional
SPD matrix gives $96\times97/2=4,656$ coordinates. The basis removes channel
redundancy but is not a universal atlas: it is re-estimated for every fold and
dataset. Transfer therefore means that the recipe works on a new cohort, not
that one classifier or one EOF orientation is transported between montages.
Because PCA orders total variance rather than discriminative variance, strong
artefacts can also consume part of the 16-mode budget.

\paragraph{Density tangent: modal-compressed delay covariance.}
Signals are projected onto the first 12 non-constant eigenmodes of a fixed
cotangent scalp Laplacian, then stacked at the same six delays. For OAS state
covariance $C_X$, we separate scale and shape as
\begin{equation}
  \rho_X=\frac{C_X}{\operatorname{tr}(C_X)},\qquad
  s_X=\sqrt{72}\log\!\left(\frac{\operatorname{tr}(C_X)}{72}\right).
\end{equation}
The trace-one operator $\rho_X$ is mapped with the standard AIRM tangent map;
$s_X$ is appended as one explicit coordinate. The representation therefore
retains the full operator rather than a handful of quantum-inspired scalar
observables.

\featureanatomy The mesh projection has 12 modes and six delays, so $C_X$ is a
$72\times72$ matrix and its tangent coordinates have dimension 2,628; the scale
coordinate produces 2,629 features in total. Trace normalisation has a precise
amplitude effect. Under a common trial rescaling $X\mapsto aX$, $\rho_X$ is
unchanged, whereas $s_X$ shifts by $2\sqrt{72}\log|a|$. The complete map is
therefore not scale invariant: nuisance gain is isolated in one direction
rather than deleted. The factor $\sqrt{72}$ is consistent with the Frobenius
length of an isotropic log-scale change in a 72-dimensional SPD space. Shape
coordinates are referenced to a training-fold geometric mean and are passed to
the classifier without an additional empirical rescaling, preserving that
analytic balance between shape and scale.

\paragraph{Root density.}
Using the same mesh--delay state, we eigendecompose $\rho_X=U\Lambda U^\top$
and vectorise its positive square root $U\Lambda^{1/2}U^\top$ with
$\sqrt{2}$-weighted off-diagonal entries. Log total scale is appended. This is
a fixed square-root feature embedding; because no pair-dependent unitary
alignment is performed, we do not identify it with an exact Bures distance.

\featureanatomy The state covariance and trace decomposition are identical to
density tangent, allowing a focused comparison of coordinate maps. Eigenvalues
of $\rho_X$ are floored at $10^{-12}$ before taking square roots, which avoids
numerical instability without changing well-resolved modes. The weighted upper
triangle again contains 2,628 values and $\log[\operatorname{tr}(C_X)/72]$
adds the 2,629th. Since
$\|\rho_X^{1/2}\|_F^2=\operatorname{tr}(\rho_X)=1$, the shape block lies on a
unit Frobenius sphere and a linear model measures chord-like differences in
square-root amplitude. Unlike density tangent, no training-fold SPD reference
is required; a conventional training-fold standardisation is applied to the
final coordinates. This simpler embedding is computationally cheaper but does
not preserve the affine-invariant local geometry of the density manifold.

\paragraph{Complex EOF--Hankel.}
Twelve training-fold EOF scores are converted to analytic signals and stacked
at delays $\{0,2,4\}$. Concatenating real and imaginary parts yields a
72-dimensional state. Its OAS covariance and training-fold tangent coordinates
encode amplitude and quadrature propagation jointly.

\featureanatomy The analytic signal is computed after the common 8--32~Hz
preprocessing, not separately in the three internal bands. For EOF score
$u_k(t)$, the Hilbert transform supplies a quadrature component
$\mathcal{H}u_k(t)$; concatenating real and imaginary parts over three delays
creates $2\times12\times3=72$ state variables. The covariance therefore
contains real--real, quadrature--quadrature, and cross-quadrature blocks rather
than only a complex coherence magnitude. Its weighted upper triangle has 2,628
coordinates. As with extended EOF, the PCA basis and tangent reference are
fitted without held-out trials. Analytic construction can be sensitive to epoch
boundaries and to small timing changes, so the added phase information is also
a plausible source of variance rather than an unqualified enrichment.

\paragraph{Seismic multitaper polarization.}
Each epoch is projected onto 12 fixed scalp eigenmodes. Four discrete prolate
Slepian tapers with time-bandwidth product 2.5 estimate a complex spectral
density matrix in each band. After trace-proportional ridge regularisation, a
Hermitian matrix $H=A+iB$ is realified as
\begin{equation}
  \mathcal{R}(H)=\begin{bmatrix}A&-B\\B&A\end{bmatrix},
\end{equation}
and mapped to band-specific AIRM tangent coordinates. Unlike scalar
rectilinearity indices, this keeps the complete polarization matrix.

\featureanatomy For each taper, the 12 modal signals are Fourier transformed.
All taper--frequency vectors whose Fourier bins fall inside one of the three
bands are pooled to estimate
$H_b=S_b^{-1}\sum_{s=1}^{S_b}v_s v_s^{\ast}$. A ridge equal to
$10^{-3}\operatorname{tr}(H_b)/12$ is added to its diagonal. Realification
turns each $12\times12$ Hermitian matrix into a $24\times24$ real SPD matrix,
whose tangent vector has $24\times25/2=300$ coordinates; three bands give 900.
The construction retains signed quadrature relations through the imaginary
part of $H_b$, but averages over frequency bins within each band. It therefore
trades fine spectral location for a lower-variance estimate of modal
polarization. Separate training-fold tangent references prevent one band from
setting the scale of another.

\paragraph{Modal log-covariance.}
Band-limited signals are projected onto the first 12 non-constant scalp
Laplacian modes. For each band, the trace-normalised OAS covariance is mapped
by the matrix logarithm at the identity and its weighted upper triangle is
vectorised. Centred log total energies across bands are appended. This is a
fixed modal feature map, not a learned tangent reference.

\featureanatomy A fourth-order zero-phase Butterworth filter is applied in each
internal band before projection. A $12\times12$ OAS covariance contributes 78
weighted upper-triangular log coordinates per band, giving 234 matrix features.
The three appended values are
\begin{equation}
  r_b=\log e_b-\frac{1}{3}\sum_{b'=1}^{3}\log e_{b'},
  \qquad e_b=\operatorname{tr}(C_b),
\end{equation}
and sum to zero, so they carry two degrees of freedom although all three are
retained for a symmetric implementation. The final dimension is therefore
237. Trace normalisation deletes absolute scale inside each band, while the
$r_b$ block restores relative energy between bands. A common rescaling of the
entire epoch cancels from both blocks. The identity reference makes the map
portable across folds, at the cost of not centring it on the cohort's typical
covariance.

\paragraph{Alpha-CoDA power.}
For channel powers $p\in\mathbb{R}_{+}^{C}$, the alpha transformation with
$\alpha=1/2$ is
\begin{equation}
  z_i=\frac{C\,p_i^{\alpha}/\sum_j p_j^{\alpha}-1}{\alpha}.
\end{equation}
It represents relative spatial allocation of power and is invariant to common
positive rescaling. Coordinates from the three bands are concatenated.

\featureanatomy Each band is filtered independently and reduced to one
whole-epoch mean-square power per channel; the selected static variant has no
temporal windows. Numerical powers are floored at $10^{-30}$ before the alpha
map is evaluated in the log domain. The coordinates obey $\sum_i z_i=0$.
Although the intrinsic dimension is $C-1$, all $C$ symmetric coordinates are
retained, giving $3C$ outputs after concatenation and a rank-deficient but
well-defined input to $L_2$-regularised logistic regression. Closure is
performed separately in every band, so the representation is invariant not
only to a common trial gain but also to an arbitrary common gain applied to one
band. That is a strong nuisance removal: it also discards between-band total
power and every cross-channel covariance term. Training-fold standardisation
sets the relative weight of electrodes after this deterministic transform.

\paragraph{Cosmological scalp harmonics.}
Alpha-compositional power maps are regarded as samples of a field on a partial
sphere. Real spherical harmonics through degree $\ell=4$ are fitted by a
regularised inverse whose penalty grows with the Laplace--Beltrami eigenvalue
$\ell(\ell+1)$. Signed coefficients are retained because sagittal orientation
is discriminative for left/right imagery; one reconstruction residual per band
quantifies unmodelled partial-sky structure.

\featureanatomy Standard 10--05 electrode coordinates are normalised to the
unit sphere and evaluated in the 25 real harmonics with $0\leq\ell\leq4$. If
$Y$ is this partial-sphere design, the coefficient operator is
\begin{equation}
  P=\left(Y^\top Y+\operatorname{diag}\left\{
  10^{-2}[1+\ell(\ell+1)]^2\right\}\right)^{-1}Y^\top.
\end{equation}
The regulariser is essential on the 22-channel BNCI montage, where 25 harmonic
columns would otherwise be underdetermined. Each of the three alpha-compositional
band-power fields yields 25 signed coefficients and one root-mean-square
reconstruction residual, for $3\times26=78$ features. The map is fixed by the
montage and has no EEG-fitted parameters before standardisation. It is
deliberately not rotation invariant: a left--right reflection changes oriented
$m$ coefficients, which is precisely the information needed for handedness.
The residual prevents all high-spatial-frequency content from being silently
treated as zero, but compresses that content to one unsigned number per band.

\paragraph{Magnetic phase graph.}
The analytic phase of each channel defines the complex phase-locking matrix
\begin{equation}
  Z_{ij}=\frac{1}{T}\sum_t e^{i(\phi_i(t)-\phi_j(t))}.
\end{equation}
The representation contains per-node PLV strength, weighted sine phase flow,
mean and dispersion of edge locking, weighted absolute phase lag, and triangle
frustration. The selected \texttt{MagneticGraph\-Extras} ablation contains no
alternative graph spectrum or geometry.

\featureanatomy This candidate operates on the common broadband 8--32~Hz
epoch. The Hilbert analytic signal is normalised to unit phase before forming
$Z$, and its diagonal is set to zero. The selected block contains $C$ mean PLV
strengths and $C$ degree-normalised sine flows,
\begin{equation}
  f_i=\frac{\sum_j |Z_{ij}|\sin[\arg(Z_{ij})]}
  {\sum_j |Z_{ij}|},
\end{equation}
plus four graph summaries: mean and standard deviation of $|Z_{ij}|$, a
PLV-weighted absolute phase lag, and product-weighted triangle frustration.
The dimension is consequently $2C+4$. PLV magnitudes and closed-triangle phase
flux are invariant to independent node rephasings, whereas the signed node
flow intentionally retains a sensor-referenced orientation. All blocks are
invariant to a single common phase shift. No eigenvalues of a magnetic
Laplacian are included in this frozen ablation, despite the family name.

\paragraph{Cycle filtration.}
Within each band, every channel triangle $(i,j,k)$ receives strength
\begin{equation}
  g_{ijk}=\left|Z_{ij}Z_{jk}Z_{ki}\right|^{1/3}.
\end{equation}
At edge thresholds
$\{0,0.2,0.4,0.6\}$, we retain nodewise incident-cycle strength, active-cycle
fraction, mean, and dispersion. This is an explicit edge-strength filtration;
we do not claim that it computes persistent homology.

\featureanatomy Before constructing $Z$, each epoch is filtered separately in
the three internal bands. A triangle is active at threshold $\tau$ only when
its weakest edge satisfies
$\min(|Z_{ij}|,|Z_{jk}|,|Z_{ki}|)\geq\tau$. For every threshold, the feature
block contains the mean strength of active incident triangles for each node,
the fraction of all $\binom{C}{3}$ triangles that are active, and the global
mean and standard deviation of their strengths. Empty selections are encoded
as zero. Hence each band--threshold pair contributes $C+3$ values and the total
dimension is $3\times4(C+3)=12(C+3)$. The selected ``magnitude'' mode discards
the phase of the Wilson loop; a gauge-holonomy variant was tested during
development but is not the reported candidate. Enumerating all triangles makes
the cost cubic in channel count, explaining why montage density matters more
for this feature than for the matrix-valued candidates.

\subsection{Controls and classification}

Four controls delimit the benchmark. CSP uses eight components followed by
shrinkage LDA. The broadband Riemannian control estimates an OAS channel
covariance, maps it to the AIRM tangent space at a training-fold geometric
mean, and fits logistic regression. The spectral control repeats the covariance
and tangent mapping in the intervals 8--12, 12--16, 16--20, 20--24, 24--28,
and 28--32~Hz before concatenation. The post-hoc ACM control follows the
block-Hankel construction in the pyRiemann example but fixes six adjacent lags
rather than tuning order and lag: six aligned copies of the raw channel signal
are concatenated, followed by OAS, an AIRM tangent map fitted on the training
fold, and the same unscaled logistic regression with $C=1$. In pyRiemann 0.12,
\texttt{AugmentedDataset} is example code rather than an exported estimator.
Our pre-existing
\texttt{DelayStateCovarianceFeatures(delays=(0,\ldots,5))} implements the same
six-lag statistical object, with reversed block ordering and one additional
aligned boundary sample relative to the gallery helper. Thus this is a fixed,
delay-count-matched ACM-style control, not a reproduction of Carrara's nested
order--lag grid.

The corresponding dimensions make the
comparison concrete. CSP returns eight log-variance values; broadband
covariance returns $C(C+1)/2$ values (253, 1,953, and 2,080 in the three
montages); spectral covariance returns six times those counts; and ACM returns
8,778, 69,378, and 73,920 tangent coordinates. The controls therefore cover
both a compact supervised spatial filter and instantaneous, spectral, and
delay-augmented second-order representations.

Root density, modal log-covariance, Alpha-CoDA, spherical harmonics, phase-graph
summaries, and cycle filtration are standardised within the training fold.
Tangent coordinates from broadband covariance, ACM, spectral covariance,
extended EOF, density tangent, complex EOF, and multitaper polarization are
used in their native tangent scaling. All non-CSP pipelines use $L_2$-regularised
logistic regression with fixed $C=1$, the \texttt{lbfgs} solver, and a maximum
of 2,000 iterations. No candidate-specific classifier, penalty, dimension
reduction, or probability calibration search was performed.

\subsection{Evaluation and statistical analysis}

MOABB's within-session evaluation is applied separately to every participant
and session using five shuffled stratified folds (seed 42) and identical splits
for all pipelines. ROC--AUC is the primary metric. Fold scores are first
aggregated by MOABB; the resulting session scores are then averaged to one
value per participant and pipeline, which is the unit of inference.

Fold-local fitting prevents direct use of held-out observations, but shuffling
can place temporally adjacent trials in train and test folds. The design may
therefore be optimistic under within-session drift and does not measure
cross-session model transfer.

Within each dataset, a Friedman test first assesses the global null that all
pipelines have equal rank. Each of the ten cross-disciplinary candidates is
then compared with the broadband covariance tangent-space control by a paired
two-sided Wilcoxon signed-rank test. Holm correction is applied over the ten
candidate comparisons separately within each dataset. We report the paired
mean difference, a participant bootstrap 95\% interval with 10,000 resamples,
the median difference, and the number of participants improved/tied/degraded.
CSP and spectral covariance are reference controls and are not included in the
ten-hypothesis novelty family. The within-dataset ten-test families are the
primary multiplicity analysis. As a stricter, explicitly secondary sensitivity
analysis, we also apply one Holm correction over all 30 candidate--dataset
comparisons. Because the tests are two-sided, a corrected rejection is called
positive transfer only when the paired effect is positive; a significant
effect in the opposite direction is recorded as degradation, not transfer.
BNCI informed candidate selection and is exploratory regardless of its adjusted
value. With $n=9$, its exact two-sided Wilcoxon tests have coarse resolution
(the smallest ten-test Holm value is 0.039), and its percentile-bootstrap
intervals can undercover; neither should be read as a precise uncertainty
estimate.

The density-tangent comparison with ACM was added during manuscript audit and
is therefore secondary rather than part of the original novelty family. We
report participant-paired effects and use one Holm correction over its two
external comparisons (Lee and Physionet); BNCI is shown descriptively.

Transportability is assessed at two levels. First, we report whether the sign
of a candidate's mean effect relative to broadband covariance is consistent
over the three datasets. Second, we compare dataset-level mean ranks and their
Spearman correlations. With only three datasets, these rank correlations are
descriptive rather than population-level inference. No pooled participant test
is used because it would treat participants from different datasets as
exchangeable and allow the largest cohort to dominate.

All feature parameters, exclusions, and statistical procedures were frozen in
the analysis script. The recorded environment was Python 3.11.15, MOABB 1.5.0,
MNE 1.12.1, NumPy 2.4.6, pandas 3.0.3, pyRiemann 0.12,
scikit-learn 1.9.0, and SciPy 1.17.1; the random seed was 42. The manuscript
tables and figures are generated directly
from participant/session score files. This separation between model fitting and
report generation prevents prose-level transcription from becoming a second
source of numerical results.

\section{Results}

\subsection{Standalone decoding performance}

\begin{table*}[t]
\centering
\caption{Participant-level within-session ROC--AUC (mean $\pm$ standard deviation). Average rank gives equal weight to each dataset; lower is better. Controls are separated from the ten candidates.}
\label{tab:performance}
\small
\begin{tabular}{lrrrr}
\toprule
Representation & BNCI2014-001 & Lee2019-MI & PhysionetMI & Mean rank \\
\midrule
Density tangent & 0.912 $\pm$ 0.082 & 0.847 $\pm$ 0.116 & 0.735 $\pm$ 0.163 & 1.00 \\
Root density & 0.872 $\pm$ 0.121 & 0.828 $\pm$ 0.119 & 0.720 $\pm$ 0.171 & 3.67 \\
Extended EOF--Hankel & 0.895 $\pm$ 0.102 & 0.805 $\pm$ 0.121 & 0.711 $\pm$ 0.176 & 4.00 \\
Complex EOF--Hankel & 0.881 $\pm$ 0.108 & 0.792 $\pm$ 0.123 & 0.705 $\pm$ 0.183 & 5.33 \\
Multitaper polarization & 0.862 $\pm$ 0.122 & 0.819 $\pm$ 0.122 & 0.703 $\pm$ 0.171 & 6.00 \\
Modal log-covariance & 0.836 $\pm$ 0.130 & 0.788 $\pm$ 0.132 & 0.697 $\pm$ 0.170 & 8.00 \\
Spherical harmonics & 0.757 $\pm$ 0.126 & 0.730 $\pm$ 0.119 & 0.623 $\pm$ 0.172 & 11.33 \\
Alpha-CoDA power & 0.762 $\pm$ 0.130 & 0.728 $\pm$ 0.119 & 0.609 $\pm$ 0.153 & 11.67 \\
Phase-graph summaries & 0.741 $\pm$ 0.184 & 0.648 $\pm$ 0.110 & 0.581 $\pm$ 0.145 & 13.00 \\
Cycle filtration & 0.661 $\pm$ 0.145 & 0.606 $\pm$ 0.112 & 0.557 $\pm$ 0.129 & 14.00 \\
\midrule
Augmented covariance (ACM control) & 0.910 $\pm$ 0.090 & 0.827 $\pm$ 0.116 & 0.657 $\pm$ 0.182 & 4.67 \\
Covariance tangent (control) & 0.868 $\pm$ 0.133 & 0.834 $\pm$ 0.120 & 0.678 $\pm$ 0.192 & 5.33 \\
Spectral covariance (control) & 0.887 $\pm$ 0.113 & 0.778 $\pm$ 0.128 & 0.641 $\pm$ 0.179 & 8.00 \\
CSP (control) & 0.840 $\pm$ 0.155 & 0.781 $\pm$ 0.147 & 0.648 $\pm$ 0.178 & 9.00 \\
\bottomrule
\end{tabular}
\end{table*}

Pipeline rankings differed within every dataset (Friedman $p<10^{-9}$ in all
three cases). Density tangent had the highest participant-averaged AUC in every
cohort: 0.912 on BNCI2014-001, 0.847 on Lee2019-MI, and 0.735 on PhysionetMI.
The corresponding broadband covariance values were 0.868, 0.834, and 0.678.
Consequently density tangent was the only pipeline with mean rank 1.00 across
datasets. The next candidate mean ranks were root density (3.67), extended
EOF--Hankel (4.00), and complex EOF--Hankel (5.33), all of which retain
matrix-valued delayed, analytic, or spectral second-order structure. The
compositional, spherical-harmonic, phase-graph, and cycle summaries occupied
the lower part of the ranking.

\IfFileExists{figures/performance_heatmap.pdf}{
\begin{figure*}[t]
  \centering
  \includegraphics[width=0.94\textwidth]{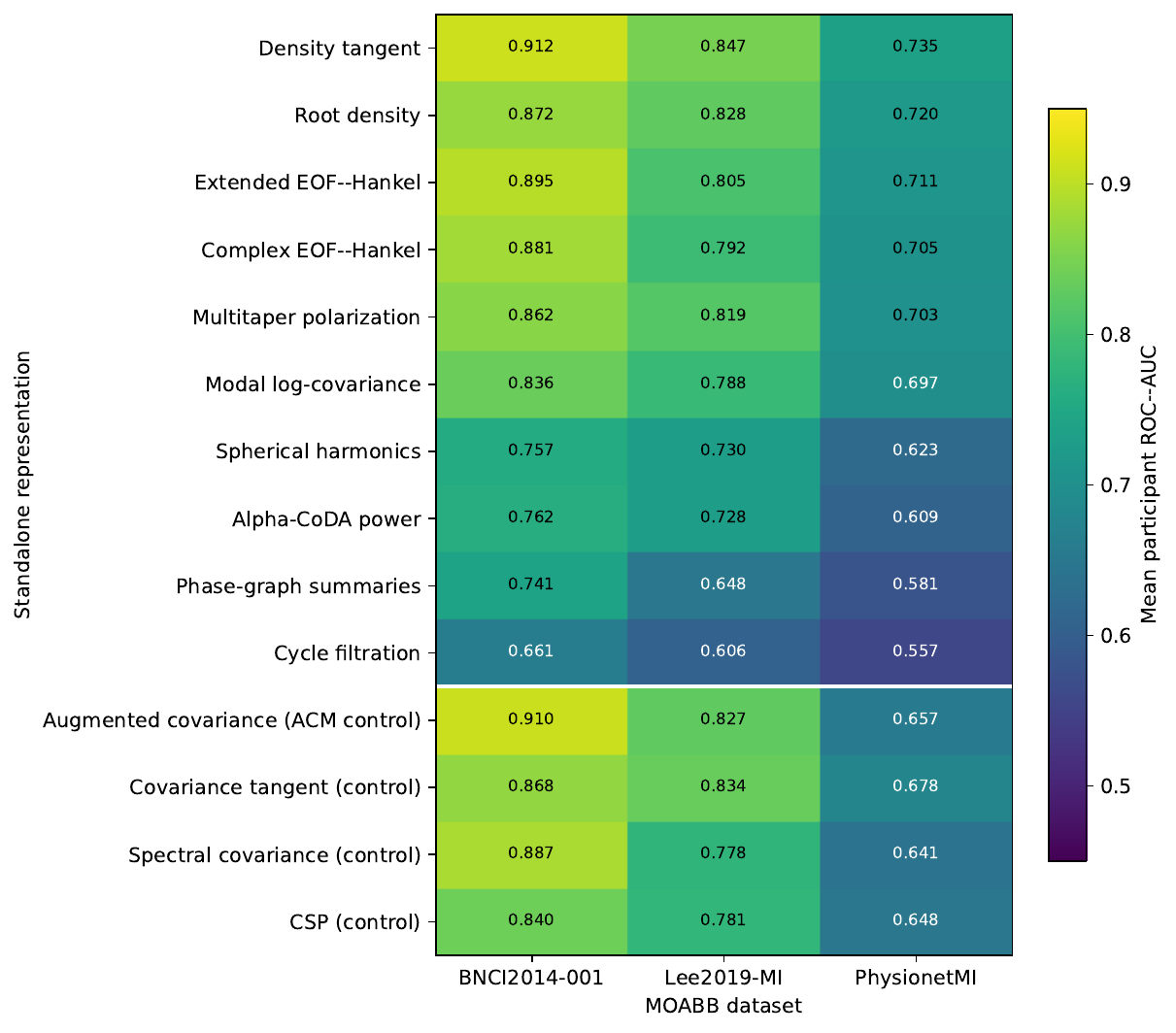}
  \caption{Participant-averaged ROC--AUC for every standalone representation
  and dataset. Rows are ordered by mean rank across datasets. Controls are
  visually separated from the ten cross-disciplinary candidates.}
  \label{fig:heatmap}
\end{figure*}}{}

\subsection{Feature-family pattern}

The ranking reveals a structured rather than monotonic effect of feature
complexity. The four leading candidates all preserve a matrix-valued delayed or
spectral state before classification. Density tangent and root density are the
most direct internal comparison because they share the same 12-mode,
six-delay, trace-normalised covariance. Density tangent exceeded root density
by 0.040 AUC on BNCI, 0.019 on Lee, and 0.015 on Physionet. These are descriptive
contrasts rather than pre-specified paired tests, but their consistent sign
suggests that the training-centred logarithmic coordinates are more useful than
the fixed square-root coordinates for this state. Root density nevertheless
ranked second on Physionet, showing that the underlying delayed modal operator
retains substantial information under both parameterisations.

The two learned-EOF states provide a second descriptive contrast. Extended EOF
used 16 real modes and six adjacent delays, whereas complex EOF used 12 analytic
modes and three spaced delays. Extended EOF was higher by 0.014, 0.013, and
0.006 AUC across the three datasets. Because mode count, delay grid, and
quadrature change together, this cannot isolate a phase penalty. It does show
that adding analytic components did not compensate for the smaller real state
under the frozen settings. Multitaper polarization, which estimates phase in
the frequency domain rather than through a Hilbert delay state, occupied a
similar middle rank. Thus complex information was decodable, but it did not
produce the cross-cohort advantage observed for scale--shape delayed
covariance.

Power-only spatial maps formed a distinct lower group. Alpha-CoDA and spherical
harmonics retain lateralised relative power but discard channel covariance and
absolute band scale. Their close performance on BNCI and Lee indicates that
compressing the compositional field into 25 partial-sphere coefficients was not
the main source of their gap from broadband covariance. The larger loss is
already present when the trial is reduced to relative channel powers. Phase
graph and cycle features were lower still, despite retaining pairwise or
higher-order synchrony. Their performance is compatible with phase relations
being less reproducible at single-trial scale under minimal artefact processing.
It does not imply that functional connectivity lacks information under other
estimators, time windows, or fusion strategies.

Output dimension alone does not order the pipelines. Extended EOF has 4,656
features and ranked near the top; density tangent has 2,629 and ranked first;
modal log-covariance has 237 and occupied the middle. The harmonic candidate has
78 features in every montage and ranked below the channel-wise Alpha-CoDA map,
whose dimension ranges from 66 to 192. Conversely, cycle filtration grows from
300 features on BNCI to 804 on Physionet but remained last. The six-band
spectral control is larger still---1,518, 11,718, and 12,480 coordinates---yet
its external means were below broadband covariance. Dimension is therefore a
relevant variance and runtime factor, but it is not a sufficient explanation of
accuracy in this benchmark.

\subsection{Paired effects against broadband covariance}

\IfFileExists{tables/paired_table.tex}{\begin{table*}[t]
\centering
\caption{Paired candidate effects relative to broadband covariance tangent space. Intervals are 95\% participant-bootstrap intervals for the mean AUC difference. $p_H$ is the two-sided Wilcoxon value after Holm correction over ten candidates within each dataset.}
\label{tab:paired}
\scriptsize
\begin{tabular}{llrrrrr}
\toprule
Dataset & Candidate & $\Delta$AUC [95\% CI] & W/T/L & Median $\Delta$ & $p$ & $p_H$ \\
\midrule
BNCI2014-001 & Density tangent & +0.044 [+0.008, +0.097] & 9/0/0 & +0.010 & 0.0039 & 0.0391 \\
BNCI2014-001 & Extended EOF--Hankel & +0.027 [+0.006, +0.060] & 8/0/1 & +0.013 & 0.0195 & 0.0977 \\
BNCI2014-001 & Complex EOF--Hankel & +0.012 [-0.011, +0.047] & 5/0/4 & +0.000 & 1.0000 & 1.0000 \\
BNCI2014-001 & Root density & +0.004 [-0.012, +0.022] & 4/0/5 & -0.004 & 0.9102 & 1.0000 \\
BNCI2014-001 & Multitaper polarization & -0.007 [-0.024, +0.010] & 4/0/5 & -0.001 & 0.5703 & 1.0000 \\
BNCI2014-001 & Modal log-covariance & -0.033 [-0.060, -0.007] & 2/0/7 & -0.022 & 0.0391 & 0.1562 \\
BNCI2014-001 & Alpha-CoDA power & -0.106 [-0.141, -0.075] & 0/0/9 & -0.094 & 0.0039 & 0.0391 \\
BNCI2014-001 & Spherical harmonics & -0.111 [-0.145, -0.079] & 0/0/9 & -0.104 & 0.0039 & 0.0391 \\
BNCI2014-001 & Phase-graph summaries & -0.128 [-0.170, -0.084] & 0/0/9 & -0.147 & 0.0039 & 0.0391 \\
BNCI2014-001 & Cycle filtration & -0.207 [-0.271, -0.139] & 0/0/9 & -0.210 & 0.0039 & 0.0391 \\
\midrule
Lee2019-MI & Density tangent & +0.013 [+0.002, +0.025] & 32/1/21 & +0.006 & 0.0216 & 0.0648 \\
Lee2019-MI & Root density & -0.006 [-0.020, +0.009] & 23/1/30 & -0.006 & 0.2055 & 0.2055 \\
Lee2019-MI & Multitaper polarization & -0.014 [-0.029, -0.001] & 21/1/32 & -0.007 & 0.0624 & 0.1248 \\
Lee2019-MI & Extended EOF--Hankel & -0.028 [-0.043, -0.014] & 15/1/38 & -0.023 & 0.0007 & 0.0030 \\
Lee2019-MI & Complex EOF--Hankel & -0.041 [-0.059, -0.024] & 15/1/38 & -0.034 & $<10^{-4}$ & 0.0004 \\
Lee2019-MI & Modal log-covariance & -0.046 [-0.063, -0.029] & 11/2/41 & -0.046 & $<10^{-4}$ & $<10^{-4}$ \\
Lee2019-MI & Spherical harmonics & -0.104 [-0.125, -0.082] & 3/1/50 & -0.111 & $<10^{-4}$ & $<10^{-4}$ \\
Lee2019-MI & Alpha-CoDA power & -0.105 [-0.127, -0.083] & 3/0/51 & -0.104 & $<10^{-4}$ & $<10^{-4}$ \\
Lee2019-MI & Phase-graph summaries & -0.185 [-0.211, -0.159] & 2/0/52 & -0.191 & $<10^{-4}$ & $<10^{-4}$ \\
Lee2019-MI & Cycle filtration & -0.227 [-0.252, -0.202] & 1/0/53 & -0.220 & $<10^{-4}$ & $<10^{-4}$ \\
\midrule
PhysionetMI & Density tangent & +0.056 [+0.037, +0.076] & 69/6/31 & +0.040 & $<10^{-4}$ & $<10^{-4}$ \\
PhysionetMI & Root density & +0.041 [+0.021, +0.063] & 62/8/36 & +0.030 & 0.0004 & 0.0023 \\
PhysionetMI & Extended EOF--Hankel & +0.033 [+0.014, +0.053] & 60/9/37 & +0.020 & 0.0029 & 0.0115 \\
PhysionetMI & Complex EOF--Hankel & +0.027 [+0.004, +0.050] & 53/5/48 & +0.005 & 0.0810 & 0.1620 \\
PhysionetMI & Multitaper polarization & +0.025 [+0.002, +0.048] & 57/4/45 & +0.020 & 0.0389 & 0.1167 \\
PhysionetMI & Modal log-covariance & +0.019 [-0.006, +0.044] & 59/4/43 & +0.020 & 0.1068 & 0.1620 \\
PhysionetMI & Spherical harmonics & -0.056 [-0.085, -0.026] & 36/1/69 & -0.055 & 0.0006 & 0.0028 \\
PhysionetMI & Alpha-CoDA power & -0.069 [-0.099, -0.039] & 34/0/72 & -0.070 & $<10^{-4}$ & 0.0004 \\
PhysionetMI & Phase-graph summaries & -0.098 [-0.132, -0.065] & 33/4/69 & -0.110 & $<10^{-4}$ & $<10^{-4}$ \\
PhysionetMI & Cycle filtration & -0.122 [-0.159, -0.085] & 29/1/76 & -0.135 & $<10^{-4}$ & $<10^{-4}$ \\
\bottomrule
\end{tabular}
\end{table*}
}{}

On the exploratory BNCI development cohort, density tangent improved all nine
participants, with mean paired
$\Delta$AUC $=+0.044$ (95\% bootstrap interval $[+0.008,+0.097]$;
$p_H=0.039$ over ten candidates). The adjusted value is at the attainable
resolution floor for $n=9$, and the percentile interval is unstable at this
sample size. On Lee the mean difference was $+0.013$
($[+0.002,+0.025]$; 32 improved, one tied, 21 degraded; $p_H=0.065$), so the
within-dataset correction left the Lee advantage uncertain. On Physionet the
difference was $+0.056$ ($[+0.037,+0.076]$; 69/6/31; $p_H<0.001$). Under the
global Holm sensitivity analysis over all 30 tests, the Physionet effect
remained significant ($p_H<0.001$), whereas the development-cohort BNCI effect
did not ($p_H=0.063$).

Two further candidates cleared multiplicity correction on the full Physionet
cohort: root density ($+0.041$, global $p_H=0.008$) and extended EOF--Hankel
($+0.033$, global $p_H=0.049$). All three surviving representations share the
delay embedding, matching the ablation (Table~\ref{tab:ablation}) in which the
delay-containing, higher-dimensional step carries the gain. Extended
EOF--Hankel was nonetheless negative on Lee, so
none of the three transferred to every cohort. Conversely, the
compositional, spherical-harmonic, phase-graph, and cycle features were
consistently below broadband covariance in mean AUC. These negative results
are part of the benchmark rather than failed auxiliary experiments.

\IfFileExists{figures/delta_rg.pdf}{
\begin{figure*}[t]
  \centering
  \includegraphics[width=0.94\textwidth]{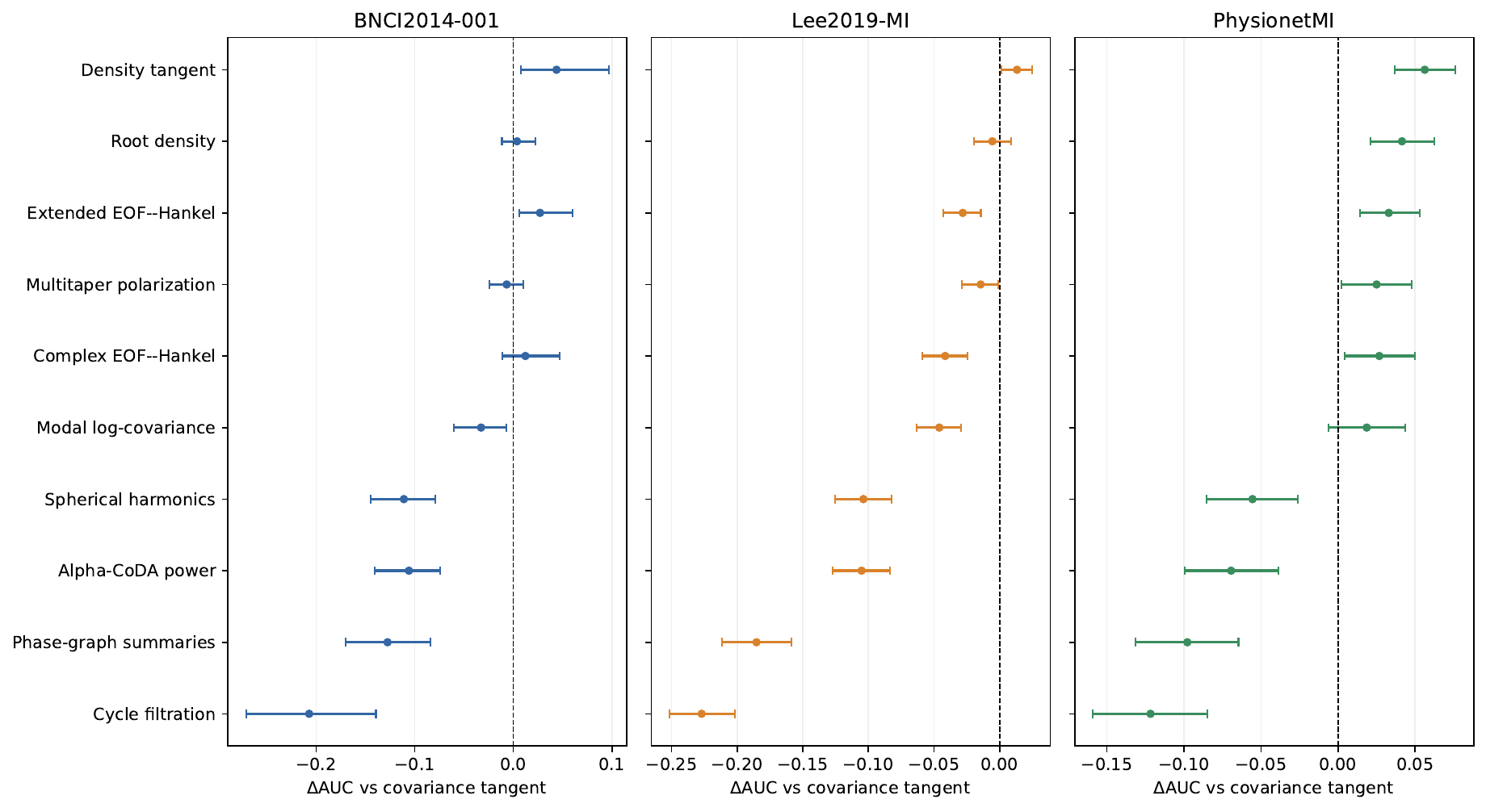}
  \caption{Paired participant-level AUC differences from the broadband
  covariance tangent-space control. Points show means and bars show 95\%
  participant-bootstrap intervals. A cross-domain representation transfers
  only if its direction and magnitude remain stable across cohorts.}
  \label{fig:deltas}
\end{figure*}}{}

\subsection{Post-hoc augmented covariance control}

\IfFileExists{tables/acm_table.tex}{\begin{table*}[t]
\centering
\caption{Post-hoc control against raw six-delay augmented covariance (ACM). $\Delta$AUC is density tangent minus ACM; intervals are participant-percentile bootstrap intervals. External $p_H$ values use one Holm correction over the Lee and Physionet comparisons. BNCI informed development and is descriptive.}
\label{tab:acm}
\small
\resizebox{\textwidth}{!}{%
\begin{tabular}{lrrrrrr}
\toprule
Dataset & ACM AUC & Density AUC & $\Delta$AUC [95\% CI] & W/T/L & $p$ & external $p_H$ \\
\midrule
BNCI2014-001 & 0.910 & 0.912 & +0.002 [-0.009, +0.017] & 4/0/5 & 0.8203 & -- \\
Lee2019-MI & 0.827 & 0.847 & +0.020 [+0.007, +0.033] & 36/2/16 & 0.0034 & 0.0034 \\
PhysionetMI & 0.657 & 0.735 & +0.077 [+0.061, +0.093] & 86/3/17 & $<10^{-4}$ & $<10^{-4}$ \\
\bottomrule
\end{tabular}}
\end{table*}
}{}

Raw six-delay ACM reached mean AUCs of 0.910, 0.827, and 0.657 on BNCI, Lee,
and Physionet. Density tangent was indistinguishable from ACM on developmental
BNCI ($\Delta=+0.002$, 95\% bootstrap interval $[-0.009,+0.017]$;
$p=0.820$). It exceeded ACM on Lee by 0.020 AUC (36 improved, two tied, 16
degraded; external Holm $p_H=0.0034$) and on Physionet by 0.077 (86/3/17;
$p_H<10^{-4}$). Thus the external advantage is not merely a re-derivation of
raw ACM. The same holds without the density normalisation: the trace-free
modal-delay covariance beat raw ACM by 0.022 on Lee (36/3/15; $p=0.0016$) and by
0.082 on Physionet (85/5/16; $p<10^{-4}$), so the gain over ACM comes from the
fixed-mode compression rather than from trace normalisation. The advantage is
not an artefact of an over-parameterised ACM: on Lee the compressed modal
covariance also beat a delay-reduced ACM at two lags (7,750 coordinates;
$\Delta=+0.014$, $p=0.018$) and three lags (17,391 coordinates; $\Delta=+0.016$,
$p=0.009$), and raw ACM accuracy fell as lags were added (0.835, 0.833, and 0.827
at two, three, and six lags). Because this control was added after the original
results were known, the comparison is a post-hoc audit result requiring
prospective replication.

\subsection{Rank stability and computational cost}

Despite differences in effect magnitude, the overall ordering was stable:
pairwise Spearman rank correlations ranged from 0.745 to 0.859. Density tangent
required median reported evaluation times of 0.77, 0.43, and 0.19 seconds per
participant/session result on BNCI, Lee, and Physionet, respectively, compared
with 0.10, 0.25, and 0.12 seconds for broadband covariance. The graph-cycle
features were both slower on the high-density montages and less accurate,
whereas the leading delayed matrix features remained computationally modest.
Raw ACM required 2.42, 12.90, and 5.73 seconds, reflecting its
montage-dependent state dimension.

The runtime pattern follows the size of the intermediate object more closely
than the final feature vector. Density tangent always reduces the montage to 12
modes before constructing a fixed $72\times72$ covariance; once that projection
has been formed, its matrix logarithm does not grow with the number of
electrodes. Extended EOF similarly fixes its state at $96\times96$, although
fitting and applying a 16-component PCA adds cost. In contrast, broadband and
six-band spectral covariance operate directly on $C\times C$ channel matrices,
and cycle filtration explicitly enumerates $\binom{C}{3}$ triangles. This
explains why cycle time increased from 0.53 seconds on the 22-channel BNCI
montage to 2.93 and 1.37 seconds on the 62- and 64-channel datasets.

These timings measure the recorded pipeline evaluation on one machine, not a
hardware-independent complexity bound. Values should be compared within
a dataset because epoch length, trial count, and caching can differ across
datasets. They nevertheless show that the leading representation is practical:
its median time remained below one second per reported participant and session
result in the frozen run, and it was faster than the six-band spectral control in
all three cohorts. The computational cost of preserving short-lag covariance
was therefore modest relative to repeating channel-space matrix estimation over
six filters.

\IfFileExists{figures/rank_stability.pdf}{
\begin{figure}[!htbp]
  \centering
  \includegraphics[width=0.78\linewidth]{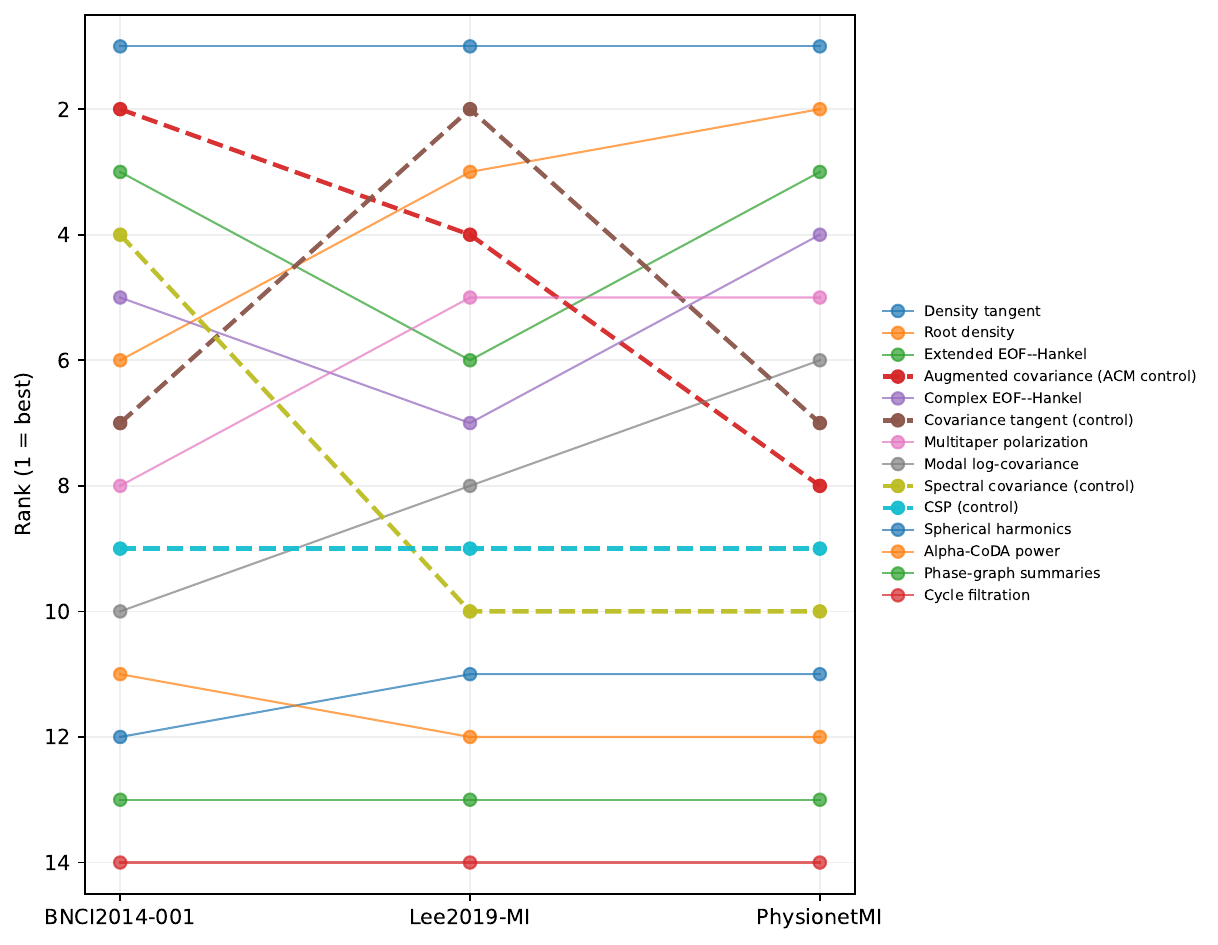}
  \caption{Dataset-specific ranks of the ten cross-disciplinary candidates
  and four controls. Parallel trajectories indicate transportability;
  crossings indicate cohort dependence.}
  \label{fig:ranks}
\end{figure}}{}

\subsection{Cross-session transfer}

Within-session evaluation can be optimistic because training and test epochs are
drawn from the same recording session. As a preliminary out-of-session check, on
the two-session datasets we trained on the first session and tested on the
second. On the nine BNCI participants the trace-free modal-delay candidate
improved mean AUC over broadband covariance by 0.041 (seven of nine
participants; paired Wilcoxon $p=0.074$) and density tangent by 0.038
($p=0.074$); on the development Lee subset (participants 6--30) density tangent
improved by 0.018 (19 of 25; $p=0.044$). These cross-session effects are smaller
than the within-session gaps and are individually only suggestive, but their
consistent sign indicates that the ranking is not purely a within-session
artefact. A full cross-session benchmark on the complete cohorts remains future
work.

\FloatBarrier
\section{Discussion}

\subsection{A qualified positive transfer result}

The strongest result is simultaneously simple and limited. Density tangent
ranked first in all three cohorts, and its gain on the cleanest external check,
Physionet, survived correction over every candidate--dataset comparison. On that
full cohort it was joined by two other delay-embedded representations, root
density and extended EOF--Hankel, which also cleared global correction; the delay
embedding they share is introduced in the only ablation step that improves all
cohorts, although that step also changes dimension. This is evidence that the
complete delayed second-order representation transfers beyond its development
dataset, with density tangent its strongest instance; it does not yet identify
lag structure as the causal component. It is not evidence of
uniform participant-level superiority: 21 of 54 Lee participants degraded and the
Lee gain did not survive within-dataset correction. Nor should the BNCI $p$ value
be read as confirmatory, because BNCI informed candidate and variant selection.
The global multiplicity analysis appropriately removes that development-cohort
claim while retaining the Physionet result.

The winning construction also does not establish a quantum advantage. The map
\begin{equation}
  C_X\mapsto(\rho_X,\operatorname{tr}C_X)
\end{equation}
is information-preserving: multiplying
the trace-one operator by its recorded scale reconstructs the delayed modal
covariance. The principal difference from the broadband control is the
statistical object: a six-delay state in fixed scalp modes with an explicit
separation of shape from scale. The present benchmark tests that complete
feature map, and the per-step ablation in Table~\ref{tab:ablation} isolates its
sequential additions: the delay-containing dimension increase carries the
accuracy gain, trace normalisation is inert, and the modal projection mainly
sets dimension. The defensible conclusion is therefore that the complete
delayed, richer second-order object transfers, not that lag structure alone or
the density-operator or quantum formalism explains motor imagery.

\subsection{A mechanistic reading of density tangent}

The leading feature differs from broadband covariance along three axes whose
individual contributions the ablation in Table~\ref{tab:ablation} separates.
First, the fixed mesh basis
reduces 22--64 electrodes to 12 smooth non-constant modes. This replaces a
montage-dependent sensor covariance with a common type of low-frequency spatial
object, although the numerical modes still differ between montages. Second,
the six-delay state adds short-lag cross-covariances. Its diagonal blocks retain
ordinary modal covariance, so the representation does not pay for dynamics by
discarding the zero-lag signal. Third, trace decomposition places all isotropic
gain in one explicit coordinate and lets the tangent map operate on relative
operator shape.

This last step can be understood without quantum terminology. An SPD covariance
contains an isotropic direction that changes all eigenvalues together and
anisotropic directions that change their ratios and eigenvectors. In a finite
sample, a large gain change can dominate numerical distances even when it is
not class specific. Normalising by the trace before estimating the tangent
reference prevents gain from moving the shape centre, while the appended scale
coordinate ensures that potentially discriminative total power remains
available. A linear classifier can then assign one coefficient to global scale
and separate coefficients to spatial--temporal shape. The improvement is
consistent with this conditioning argument, but the benchmark does not measure
condition numbers or isolate the trace operation, so it should remain a
mechanistic hypothesis.

The shared-state comparison with root density narrows the interpretation. Both
features see the same modal signals, delays, OAS covariance, and explicit total
scale. Their consistent mean difference is compatible with a role for the
treatment of shape, but it does not isolate it because density tangent uses
native tangent scaling whereas root-density coordinates are standardised before
the same $C=1$ classifier. The logarithmic tangent expands relative changes near the training-fold
centre, while the square-root map is a fixed global embedding with no
cohort-specific reference. The former may better linearise the range occupied
by motor-imagery trials. Yet a direct density-versus-root paired test was not
part of the multiplicity family, and the two branches must be rerun under a
common scaling policy before any conclusion about the superiority of one
matrix geometry is drawn.

Finally, the large Physionet gain should not be interpreted as proof that one
component specifically solves high-density EEG. Physionet also differs in trial
count, session structure, recording protocol, and participant population. The
result shows that the complete representation was robust to that bundle of
changes. The sequential ablation in Table~\ref{tab:ablation}, evaluated on the
full cohorts, associates this robustness with the delay-containing
dimension increase and, at 64 channels, with the modal reduction, rather than
with the trace-based scale-shape separation.

\subsection{What transfers is an invariance, not a disciplinary label}

The benchmark was designed to distinguish an appealing analogy from a useful
inductive bias. A method's scientific origin cannot by itself predict EEG
performance. Extended EOFs, density operators, polarization matrices, and
spherical multipoles differ because they preserve different statistics: delayed
second-order dependence, operator shape, complex cross-spectral phase, or
large-scale spatial orientation. The relevant explanatory axis is therefore
the retained information and invariance, not whether the vocabulary is
meteorological, quantum, seismic, or cosmological.

The final three-dataset results should be interpreted accordingly. A positive
effect on BNCI alone is evidence that the feature can encode motor imagery, but
not that the representation transfers. A stable effect on Lee and Physionet,
whose montages and acquisition protocols differ substantially, is the stronger
criterion. Conversely, a candidate near chance after removal of absolute power
is informative about the task: it suggests that its intended higher-order or
phase-specific structure is not stable enough to support left/right decoding
under the present sample sizes.

\subsection{Negative features define what cannot be removed}

The lower-ranked candidates are useful because each makes a stronger invariance
assumption than broadband covariance. Alpha-CoDA removes total power separately
inside every band and retains no cross-channel products. Spherical harmonics
add spatial smoothness and compression to that same scale-free field. Their
consistent deficit indicates that relative lateralised power alone is not an
adequate replacement for the full covariance object under this protocol.
Absolute scale, covariance interactions, or both carry information that the
composition intentionally removes. This is a boundary condition on a standalone
feature, not an argument against using compositional coordinates as an
additional robust branch.

The phase features remove a different part of the signal. Phase-graph summaries
normalise every analytic sample to unit magnitude before estimating PLV, so
event-related power modulation is absent by construction. Cycle filtration then
compresses the edge matrix into thresholded triangle magnitudes and, in the
reported variant, discards loop phase. These transformations ask a difficult
single-trial question: is coordination reproducible after amplitude is removed?
The answer was negative in mean AUC across all three datasets. Because no
artefact rejection or time-resolved phase segmentation was used, the result is
specific to whole-epoch PLV and should not be generalised to all connectivity
methods.

The harmonic result also clarifies what montage portability does and does not
mean. A fixed low-degree basis provides comparable spatial semantics---smooth,
lateral, anterior--posterior, and higher multipoles---but it cannot recover
information that was discarded before projection. A representation may have an
elegant coordinate system and still be statistically insufficient for the
task. Conversely, the learned EOF basis has no cross-dataset alignment but can
perform well because each cohort supplies a useful variance-adapted state. In
this benchmark, portability of the fitting rule mattered more than literal
identity of feature axes.

\subsection{Why the matrix-valued candidates are strong controls on the claim}

Several candidates ultimately return tangent coordinates of SPD matrices. This
does not make the benchmark a comparison of geometries. The metric is held
fixed; the statistical object changes. Raw channel covariance, lagged EOF
covariance, trace-one delayed modal covariance, complex analytic covariance,
and multitaper polarization encode different sufficient statistics. Holding
the downstream AIRM map fixed is precisely what isolates the object. In
contrast, Bures--Wasserstein, log-Euclidean, BKM, Grassmann, symplectic, and
ultrametric variants were excluded because they change the geometry and belong
to a different research question.

The two quantum-inspired candidates also illustrate the limit of disciplinary
language. Both operate on ordinary classical covariance estimates. Calling a
trace-one SPD matrix a density operator is mathematically valid, but it creates
scientific value only if the scale--shape separation or square-root coordinates
generalise. We therefore avoid the terms quantum computation and quantum
advantage throughout.

\subsection{Implications for feature engineering}

Two practical principles follow from the full ranking. First, enrich a strong
sufficient statistic before replacing it. The successful candidates preserve
zero-lag covariance and add delay, analytic, or spectral blocks; the weakest
candidates replace power covariance with aggressively normalised summaries.
For motor imagery, a new invariance should therefore be tested both as a
standalone object and as a controlled extension of covariance before being used
in a fusion model.

Second, distinguish a fixed spatial prior from a fixed numerical basis. Mesh
eigenmodes provide the same construction across montages, but their samples and
eigenvectors change with electrode positions. They regularise spatial scale
without pretending that a coefficient learned on one montage can be copied to
another. EOFs make the opposite trade: the basis is entirely data adaptive,
but fold-wise fitting keeps the procedure leakage safe. Both can transfer as
algorithms even though neither supplies identical columns across datasets.

\subsection{Relation to our earlier benchmark}

Our earlier study asked whether one established decoder was optimal for most
participants under favourable within-session conditions and found substantial
participant-level heterogeneity \citep{vasques2026rankings}. The present work
does not repeat its 1,056-configuration grid. Instead, the earlier finding that
representation matters more than downstream scaler or classifier motivates us
to hold the decoder nearly fixed and broaden the representation space. The two
papers therefore address consecutive questions: first, whether a universal
pipeline exists within the usual vocabulary; second, whether external
scientific vocabularies offer representations whose value survives a change of
dataset.

\subsection{Limitations}

Several limitations bound these claims. The candidate inventory is broad but not
exhaustive, and a negative result for one adaptation is not evidence against its
discipline. BNCI informed development, so its scores are exploratory, and related
matrix candidates were inspected on part of Lee, so Lee is not a fully untouched
holdout. Only three datasets are used, now at full cohort size (169
participants), but generalisation across three cohorts is less certain than
participant-level uncertainty. Within-session evaluation measures
transportability of performance across cohorts, not model transfer; shuffled
folds can be optimistic under session drift, and cross-session transfer remains a
separate experiment.

A dimension-matched surrogate on Lee clarifies why the delayed object helps: at a
fixed 60-dimensional state, replacing the cross-lag structure with additional
spatial modes left accuracy unchanged (cross-lag 0.848 versus spatial 0.844;
$p=0.91$), so the gain reflects the added second-order dimension rather than
cross-lag dynamics specifically. One caveat remains on the object: the ACM
control of \citet{carrara2023augmented} was added during audit, so although
density tangent exceeded it on Lee and Physionet under the same split and
classifier, this post-hoc comparison is not a prospectively frozen confirmation. Minimal preprocessing may
also penalise phase- and topology-based candidates, the participant bootstrap is
fragile on BNCI ($n=9$), and several constructions reuse established blocks (PCA,
OAS, PLV, AIRM), so we claim novel adaptations and a systematic benchmark, not
invention of those components.

\section{Conclusion}

This work tests whether hand-designed representations from other scientific
fields survive contact with motor-imagery EEG. Ten standalone candidates were
implemented under a common within-session MOABB protocol with fold-local fitting
and evaluated against
four established controls on three datasets. By excluding feature fusion and
metric shopping, the benchmark attributes success or failure to one statistical
representation at a time. Its central contribution is therefore an empirical
map of which imported invariances transfer, which remain dataset-specific, and
which discard the second-order power structure that dominates left/right motor
imagery. Density tangent ranked first in all three cohorts and, on the full Physionet
cohort, was the strongest of three delay-embedded representations to clear
global correction on an external dataset; its Lee effect nevertheless remained
indistinguishable from zero after correction. The most robust finding, however,
is the ACM comparison: a raw six-delay augmented covariance collapses on the
64-channel montage, whereas the same object compressed onto twelve fixed scalp
modes ranks first and beats raw ACM by up to 0.077 AUC. A per-step ablation
attributes this to the delay embedding and its modal compression, not to the
trace normalisation, so the trace-free modal delayed covariance is the
actionable representation and ``density tangent'' is its equivalent,
information-preserving variant. The transferable ingredient is a fixed modal
compression that makes a delay-augmented covariance estimable when trials are
scarce, whereas disciplinary labels alone do not make a feature transfer.

\section*{Data availability}

All datasets are publicly available through MOABB. The exact datasets,
participant identifiers, task restriction, and preprocessing used in the
benchmark are specified in the Methods. No new raw EEG data were generated.

\bibliographystyle{unsrtnat}
\bibliography{references}

\section*{Competing interests}

O.O. and P.B. are co-founders of Inclusive Brains, which maintains a research
partnership with IBM on artificial intelligence, neurotechnologies, and quantum
computing; this partnership did not influence the design, analysis, or
interpretation of this study. X.V. is Vice-President and Chief Technology
Officer of IBM Technology France and Head of Laboratory at the Institut du
Neurone. O.O. is Professor at MBZUAI,
CEO of Inclusive Brains, Chairman of the Biotech Dental Group AI Institute, and
a shareholder and former President of EMOTIV Inc. No other competing interests
are declared.

\end{document}